\documentstyle[tighten,prl,epsf,floats,aps]{revtex} 

\def\comment#1{}

\begin{document}

\twocolumn[
\hsize\textwidth\columnwidth\hsize\csname @twocolumnfalse\endcsname

\date{\today, submitted to Heraklion Conference of unconventional 
superconductors.}
\draft
\preprint{\today}

\title{Infrared properties of exotic superconductors.}

\author{T.~Timusk } 

\address{Department of Physics and Astronomy, 
McMaster University, Hamilton, Ontario, Canada L8S 4M1}

\maketitle

\begin{abstract}
{The infrared spectra of the non-traditional superconductors share certain
common features. The lack of a gap signature at $2\Delta$ and the residual 
conductivity are the consequence of a d-wave order parameter. 
The high $T_c$ 
materials, the organic conductors and the heavy Fermion materials have a strong 
mid-infrared absorption band which can be interpreted as strong coupling of 
the carriers to electronic degrees of freedom which leads to a breakdown of 
the  Fermi liquid picture. The cuprates and the organic charge transfer 
salts are unique in possessing an intrinsic low dimensionality. The charge 
transport normal to the highly conducting direction is incoherent down to 
the lowest temperatures and frequencies.} 
 
\end{abstract} 
\vspace{3mm}
]

\subsection{INTRODUCTION}

\setcounter{page}{1}

The discovery of high temperature superconductivity (HTSC) by Bednorz 
and M\"uller\cite{bednorz86} raised the question of the relationship 
between these materials and other, previously known unconventional 
superconductors such as the organic conductors, the bismuthates and the 
heavy Fermion materials. Since 1987 other novel superconductors such 
as the borocarbides\cite{boro} and the alkali 
fullerines\cite{hebard91} have been 
discovered. The aim of this review is to examine, from the point of view of 
one experimental technique, infrared spectroscopy, several of  these 
families of unconventional superconductors.   

Optical spectroscopy has several advantages over other techniques for the 
analysis of the transport properties of a broad range of conducting systems. 
First, it is not surface sensitive since electromagnetic radiation penetrates a 
hundreds of nanometers into the crystal. Second, relatively small lateral 
sample dimensions of the order of 500 $\mu$m are enough to yield high 
quality spectra in the interesting 200  -- 1000 cm$^{-1}$ frequency  region, 
a region relevant to the excitations in most superconductors. In contrast, 
more powerful techniques such as angle resolved photoemission or vacuum 
tunneling demand ultrahigh-vacuum cleaved, virgin surfaces and as a result 
have been applied to only a few systems. Magnetic neutron scattering, 
another powerful technique, due to the weak interaction of the neutron with 
matter, demands large centimeter size crystals which are simply not 
available for most new materials. As a result, optical spectroscopy has 
become the spectroscopic technique of choice for the 
investigation of the low 
lying excitations of a large range of new materials and doping levels. 

Reflectance spectroscopy, combined with Kramers Kronig analysis, yields the 
real and imaginary parts of the optical conductivity. In systems 
with s-wave, dirty limit superconductivity, optical methods were first used to 
study the energy gap\cite{tinkham57} and later the spectrum of 
excitations.\cite{joyce70,farnworth74,farnworth76prb}. It can also be used to determine the 
superconducting penetration depth tensor. And unlike the dc conductivity which 
becomes infinite in the superconducting state and shorts out all parallel 
channels of conductivity, infrared techniques can be used to 
measure the optical conductivity of quasiparticles below 
$T_c$.\cite{romero92} Even better results 
are obtained with the closely related technique, microwave 
absorption.\cite{bonn92} 

In what follows we review the optical properties in the infrared of a range of 
unconventional superconductors as revealed by reflectance spectroscopy. We 
start by setting the stage with the discussion of conventional BCS 
superconductors and then move to several families of less and less 
conventional superconductors starting with materials that seem to be almost 
conventional but show clear signatures of deviation from familiar behaviour.

There is a grey area in all classification schemes. For unconventional 
superconductors it includes 
materials that are distinguished by high transition temperatures and where 
relatively little reliable infrared data exists. This includes materials such 
as the A15 compounds,\cite{farnworth_thesis} the borocarbides, 
and the doped fullerines. Some of these materials are discussed 
by other contributors to this conference. 

\subsection{BCS SUPERCONDUCTORS IN THE INFRARED}

\begin{figure}[t]
\leavevmode
\epsfxsize=\columnwidth
\centerline{\epsffile{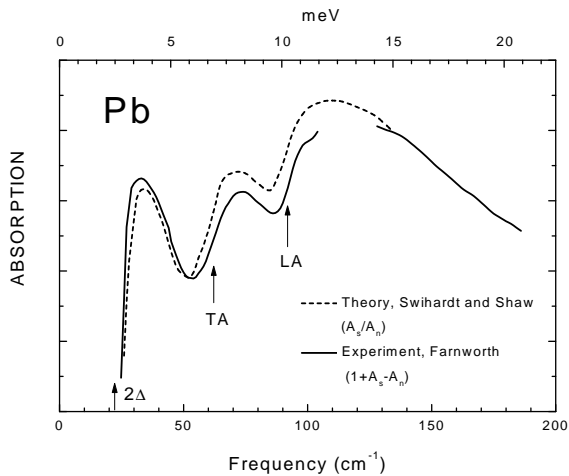}}
\caption{Infrared absorption of lead in the superconducting state divided by 
the normal state absorption (arbitrary units).  This conventional BCS 
superconductor does not absorb energy below $2\Delta$ the superconducting 
gap and there is a sharp threshold of absorption at this frequency. There is a 
further increase in absorption at $h\nu = 2\Delta + \Omega_{TA}$ the onset 
of transverse acoustic phonon scattering and another one at the longitudinal 
frequency. The dashed curve is a theoretical one based on Eliashberg theory. 
There is excellent agreement between the theory and the experiment.
}
\end{figure}

One of the pioneering experiments in  
conventional superconductors was the measurement of the energy gap by far 
infrared reflectance spectroscopy by Tinkham and his 
collaborators\cite{tinkham57}.  The gap in the 
excitation spectrum of magnitude $2\Delta$ leads to a 
region of unit reflectance for $h\nu<2\Delta$. Above the gap frequency 
absorption sets in and the reflectance drops to the normal state value. 
Reflectance experiments in conventional metals are difficult since, in a 
good metal, the normal state absorption is very weak, typically $1-R=0.005$. 
To overcome the weak absorption, 
sensitive calorimetric techniques\cite{joyce70} or multiple reflection 
cavities\cite{farnworth_thesis} are used. The disadvantage of 
these methods is the 
difficulty of getting an absolute calibration.

The second important early contribution of infrared spectroscopy was the 
discovery of Joyce and Richards of phonon structure resulting from 
strong coupling effects, similar to the structure seen in  tunnelling 
spectra.\cite{joyce70} Fig 1. shows, from the work of 
Farnworth,\cite{farnworth_thesis} the difference in the absorption of a 
lead film between the superconducting and normal states. 
A magnetic field was used to destroy the superconductivity. The rise in 
absorption at 22 cm$^{-1}$  is due to the onset of normal state 
absorption at the gap energy at $h\nu=2\Delta$.  The double-peak 
structure above the gap energy is due the interaction of the superconducting 
carriers with transverse and longitudinal acoustic phonons. The dashed curve 
is a theoretical  calculation by Swihart and Shaw\cite{swihart} of the 
absorption ratio using a full Eliashberg theory including vertex 
corrections. The phonon spectrum of lead as determined by neutron scattering 
was used. It is clear from these data that it is possible to identify all 
the important features of a BCS superconductor with infrared reflectance 
spectroscopy: an infrared signature of BCS superconductivity is a region of 
unit reflectance that occurs below $T_c$ followed by phonon structure that 
reflects the electron phonon density of states which is directly responsible 
for superconducting pairing. 

As we will see in what follows, the unconventional superconductors do not 
show this simple behavior, but even in BCS superconductors there are many 
complications that go beyond the simple strong coupling behavior shown in 
Fig 1. For example, it is not always possible to see a gap signature and a 
phonon spectrum in the same sample. The onset of absorption at the gap 
frequency is due to the presence of impurities that allow for momentum 
conservation. In a clean superconductor absorption does not start at 
$2\Delta$ but at frequencies 
where inelastic processes such as phonons can take up the momentum of 
the photon. Lead films such as those shown in Fig. 1 have some impurities 
that allow for a gap to be seen but not enough to dominate the scattering 
processes and suppress the electron phonon scattering.\cite{allen71} 
A gap signature and phonons appear together in 3D metals where 
surface scattering processes result in apparent dirty limit 
behavior in pure crystals.

Furthermore, in pure 3D metals, there are surface scattering processes that 
result in apparent dirty limit behavior and allow a gap signature and 
phonons to appear in the same sample as shown in Fig. 1. 

\subsection{BKBO AN UNCONVENTIONAL OXIDE SUPERCONDUCTOR.}   
                                                                        
Ba$_{1-x}$K$_x$BiO$_3$ (BKBO), along with its cousin 
BaPb$_x$Bi$_{1-x}$O$_3$, form a family of superconductors with 
cubic perovskite structure in 
the superconducting phase where $x\approx 0.3$.\cite{mattheiss88b} The 
transition temperature of BKBO is quite high (31 K) in relation to its low 
density of states at the Fermi level, and it has been suggested\cite{batlogg89} 
that the material is closely related to cuprate high $T_c$ superconductors. 
However the optical properties suggest that they are in fact closer to the 
conventional BCS superconductors than to the exotic cuprates. 

\begin{figure}[t]
\leavevmode
\epsfxsize=\columnwidth
\centerline{\epsffile{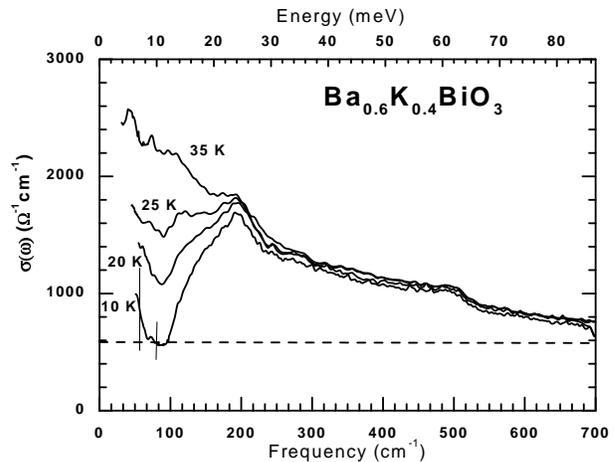}}
\caption{The optical conductivity of BKBO at several temperatures. In the 
normal state (top curve) the conductivity has a Drude peak which has a 
temperature independent width, suggesting that the scattering is elastic due 
to impurities. As the temperature is lowered below the superconducting 
transition temperature a gap like depression develops in the conductivity. We 
associate this with an s-wave superconducting gap.  There is no evidence of 
the phonon structure in the conductivity expected for conventional 
electron-phonon interaction.
}
\end{figure}

Fig. 2 shows the optical conductivity at low frequency, at four 
temperatures, for Ba$_{0.6}$K$_{0.4}$BiO$_3$ with $T_c=31 $ K from the work of 
Puchkov {\it et al.} \cite{puchkov94a} Apart from a temperature independent 
background, indicated by a dashed line, the temperature behavior is that of 
a conventional BCS, slightly dirty, superconductor. The normal state is well 
fit by a Drude peak with a width of $\approx 230 $ cm$^{-1}$ plus some direct 
transverse optical phonon absorption lines at about 200 and 500 cm$^{-1}$. A 
gap-like depression develops precisely at $T_c$ and a gap {\it opens} as 
the temperature is reduced. The superconducting gap is 
$2\Delta=90\pm10$ cm$^{-1}$ and as $T_c$ is varied with doping, the gap to 
$T_c$ ratio is constant at $2\Delta/k_BT_c = 4.2\pm 0.3$. This simple 
behavior should be contrasted with what is seen in the cuprates: there 
the gap develops already in the {\it normal state} 
(except in the limit of large overdoping), there is no $2\Delta$ gap 
signature and the gap ratio is not constant---in the underdoped range the 
gap {\it decreases} as $T_c$ is increased  with doping.\cite{renner98}

In BKBO, at 10 K, the conductivity below the gap frequency drops to a 
background conductivity shown as a  dashed line. This background absorption 
level has been obtained by extrapolation from high frequencies. Puchkov 
{\it et al.} interpret the data in terms of two channels of  conductivity, a 
Drude channel that undergoes a conventional superconducting transition and a 
second, possibly polaronic, channel that remains non-superconducting and is 
responsible for the broad temperature independent background. The two 
channel picture has also been advanced to explain the temperature dependence 
of the dc resistivity.\cite{hellman} 
Thus the infrared conductivity of BKBO is that  of  dirty s-wave 
superconductor, a behavior that is quite different from what is seen in a 
dirty d-wave superconductor as we will see below. 

Despite the development of a conventional gap structure below $T_c$, the 
BKBO spectrum does not show all the signatures of a conventional BCS phonon 
coupled superconductor, which in view of the large transition temperature is 
expected to be a strong coupling  system.  A detailed calculation, based on 
Eliashberg theory with phonons by Marsiglio {\it et al.}, bears this out 
yielding electron-phonons coupling $ \lambda=1$  and a transition 
temperature of 31 K.\cite{marsiglio96} As Fig. 3 shows, the imaginary part of 
the optical conductivity cannot be fit with this large value of $\lambda$ 
but is fit well with $\lambda=0.2$, a number that is clearly inconsistent 
with the large transition temperature and the assumed phonon spectrum taken 
from tunneling.\cite{zasadzinski91} Sharif {\it et al.} have found a similar 
low value of $\lambda$ based on tunneling measurements. They also conclude 
that BKBO is not an electron-phonon superconductor.\cite{sharif91} This weak 
coupling to phonons is consistent with the absence of Holstein phonon 
structure below $T_c$ and as Fig. 2 shows there 
is no evidence of such structure in BKBO. 

\begin{figure}[t]
\leavevmode
\epsfxsize=\columnwidth
\centerline{\epsffile{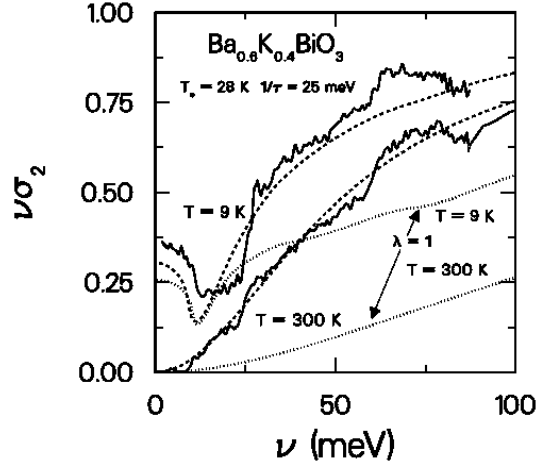}}
 \caption{The real part of the conductivity of BKBO (solid curves) and a 
theoretical calculation of the same quantity using Eliashberg theory with 
the electron-phonon interaction. There is a discrepancy between the coupling 
constant $\lambda=0.2$  (dashes) needed to fit the infrared experiments, and 
$\lambda=1$ needed to fit the high transition temperature.}
\end{figure}

\subsection{{\rm Sr$_2$RuO$_4$} and {\rm URu$_2$Si$_2$} TWO FERMI LIQUID SUPERCONDUCTORS }

In this section we discuss two materials that, at first sight appear 
unrelated, URu$_2$Si$_2$ a typical intermetallic compound showing heavy 
Fermion behavior at low temperature and Sr$_2$RuO$_4$ an oxide of Ru that is 
isostructural with the high $T_c$ superconductor La$_{2-x}$Sr$_x$CuO$_4$ 
with the ruthenium 
replacing copper. However, a closer examination shows a surprising number of 
common elements of the two ruthenium containing materials. Both materials are 
superconductors with a rather low $T_c$ of the order of 1 K. Infrared 
spectroscopy has not been used to investigate the gap structure in these two 
materials since the gap is expected to occur in the difficult submillimeter 
region and the normal state absorption will be very weak at these 
frequencies at low temperature. However, infrared can be used to study the 
normal state transport properties.

\begin{figure}[t]
\leavevmode
\epsfxsize=\columnwidth
\centerline{\epsffile{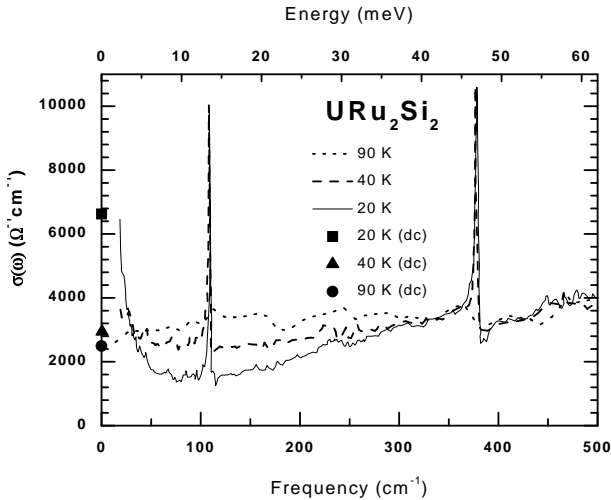}}
\caption{The normal state conductivity of URu$_2$Si$_2$. At high temperature 
the conductivity is flat and featureless due to the strong magnetic 
scattering. As the temperature is lowered below the coherence temperature 
$T_K$ a coherent Drude peak develops at low frequency. }
\end{figure}

\begin{figure}[t]
\leavevmode
\epsfxsize=\columnwidth
\centerline{\epsffile{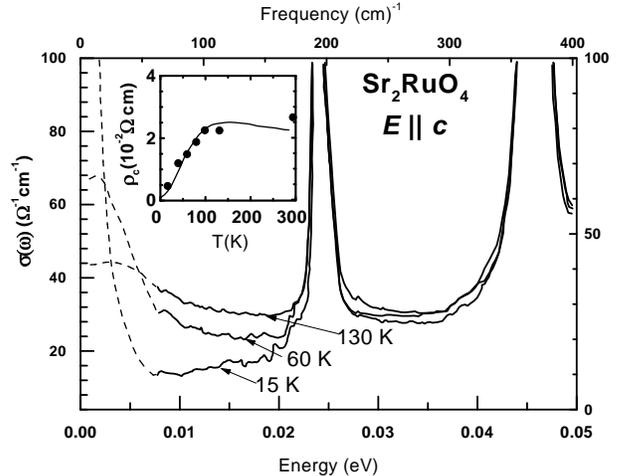}}
\caption{The normal state conductivity of Sr$_2$RuO$_4$. As in 
URu$_2$Si$_2$, at high temperature the conductivity is flat, apart from 
direct LO phonon lines. Below 100 K, a temperature where thermal fluctuation 
energy is less than the coupling energy between the planes, a coherent Drude 
peak develops at low frequency. } \end{figure}

While the inplane resistivity of Sr$_2$RuO$_4$ is metallic, normal to the 
planes the resistivity (in the c-direction) shows a "semiconducting" 
temperature dependence above 130 K and a metallic one at low 
temperatures\cite{maeno94}. Similar behavior is seen in the 
resistivity of URu$_2$Si$_2$ in all directions with a maximum at 80 K 
\cite{palstra86}. Both materials become good conductors at low temperature 
with a $T^2$ dependence of the resistivity  expected for a pure metal when the 
phonon scattering is frozen out. Superconductivity sets in at about 1 K. 

The two ruthenium materials are very different in 
their electronic properties: URu$_2$Si$_2$ gets its poor metallic behaviour 
from strong incoherent scattering which freezes out for $k_BT<k_BT_K$ where 
$k_BT_K$ is a Kondo energy scale characteristic of magnetic scattering. 
Sr$_2$RuO$_4$ on the other hand is very anisotropic in its crystal structure 
with weak coupling  between the current carrying RuO$_2$ planes. However at 
low temperature where $k_BT<t_{\perp}$ the planes couple coherently and the 
material becomes a very anisotropic 3D conductor. 

In agreement with these ideas both systems develop sharp Drude peaks at low 
temperature which are signatures of good metals. Fig. 4 shows the optical 
conductivity of URu$_2$Si$_2$.\cite{Bonn88b} At 90 K, well above the coherence 
temperature $T_K$ the 
conductivity is flat and frequency independent. As the temperature is 
lowered, a Drude peak develops at low frequency, borrowing spectral weight 
from a frequency band that is a few times $k_BT_K$. As the temperature is 
lowered further, the peak sharpens.  A detailed analysis of the spectrum in 
terms of a frequency dependent mass formalism shows that the carriers are 
massive with $m=50 m_e$ consistent with specific heat data in this heavy 
Fermion system. 

A similar onset of coherence is seen in the c-axis conductivity of the quasi 
two-dimensional material Sr$_2$RuO$_4$. Fig. 5 shows data of Katsufuji {\it 
et al.} of the c-axis conductivity.\cite{katsufuji96} The 
conductivity, apart from a few strong transverse phonon peaks, consists of a 
flat band with a slight Drude-like peak below 0.02 eV (160 cm$^{-1}$). 
Spectral weight is lost in the 100 to 200 cm$^{-1}$ region to a narrowing 
Drude peak. The author's data does not extend low enough in frequency to 
resolve the peak but the dashed lines, fit to the increasing dc conductivity 
with decreasing temperature, show the expected conductivity. As in the case 
of URu$_2$Si$_2$ the area under the Drude peak is a small fraction of the 
total infrared spectral weight suggesting a large mass for c-axis motion. 
This is verified by magneto oscillation experiments\cite{mackenzie96}. 

The two ruthenium containing systems behave in a very conventional way at low 
temperature, in that they are 3-D Fermi liquids, but at high temperature one 
shows effects of magnetism and the other of highly anisotropic transport 
properties. As we will see in the next two sections, combining 
both ingredients results in truly exotic behavior of the 
cuprates and the organic conductors.

\subsection{THE CUPRATES}

While the high superconducting transition temperature is  the defining 
property of the HTSC cuprates, several properties set the cuprate family 
apart from the conventional BCS superconductors. These include the 
anomalous normal state resistivity as emphasized early by 
Anderson,\cite{anderson88prl} the d-wave superconducting 
order parameter, the magnetism, the two-dimensional transport, and finally the 
presence of a normal state pseudogap. 

There is an extensive literature on the optical conductivity of the 
cuprates which has been reviewed recently by Basov and Timusk.\cite{basov99} 
Here we focus on those features of the optical conductivity that make the 
cuprates unique: the lack of the $2\Delta$ gap signature in the optical 
conductivity, the crucial role played by the two-dimensional nature of the 
transport and finally the the normal state pseudogap.

The absorption and the optical conductivity of a typical HTSC cuprate, 
YBa$_2$Cu$_3$O$_{7-\delta}$, is shown in Fig. 6. The normal state shows a 
Drude-like peak. Below $T_c$ the spectral weight of the peak is transferred 
to the condensate delta function at $\omega=0$ but there is no signature of 
a superconducting gap as seen in BKBO. Instead, there remains a weak 
continuous conductivity at all frequencies including in the region where one 
expects to see the superconducting gap. Careful study of the optical 
conductivity by various techniques shows that the width of the Drude peak 
diminishes by several orders of magnitude just below 
$T_c$\cite{nuss91,romero92,bonn92,hosseini99} which suggests that a gap is 
formed in the spectrum of excitations responsible for the scattering of the 
carriers. This behaviour is in striking contrast to electron-phonon 
superconductors where there are no major changes to the phonon spectrum at 
the superconducting transition. In the cuprates the excitations are 
electronic and as a gap develops in these excitations a dramatic decrease in 
scattering takes place. The optical properties of the cuprates are those of 
a clean limit superconductor where the $2\Delta$ transitions are forbidden 
because of momentum conservation. In this respect the cuprates behave 
exactly the same as the BCS superconductors. An obvious experiment to try 
would be to introduce impurities to approach the dirty limit.  


\begin{figure}[t]
\leavevmode
\epsfxsize=\columnwidth
\centerline{\epsffile{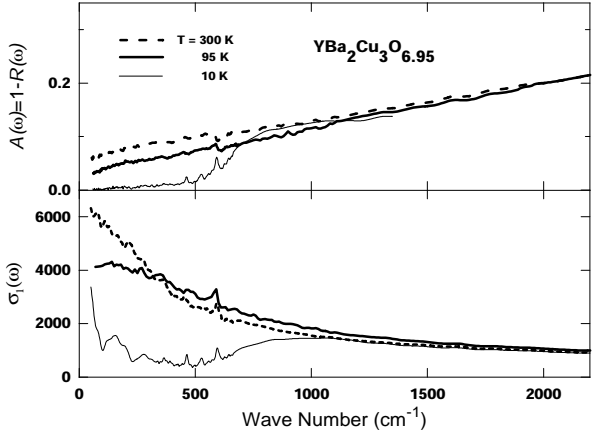}}
\caption{The absorption $A=1-R$ and the optical conductivity for optimally 
doped Y123 with $x=6.95$. A depression of A is seen below 800 
cm$^{-1}$ below the superconducting transition temperature $T_c$. 
There is a corresponding depression of conductivity in the superconducting 
state but because the system is in the clean limit with $1/\tau<2\Delta$, 
the onset of conductivity and absorption is not the superconducting gap 
which can only be seen in the dirty limit.} 
\end{figure}


\begin{figure}[t]
\leavevmode
\epsfxsize=\columnwidth
\centerline{\epsffile{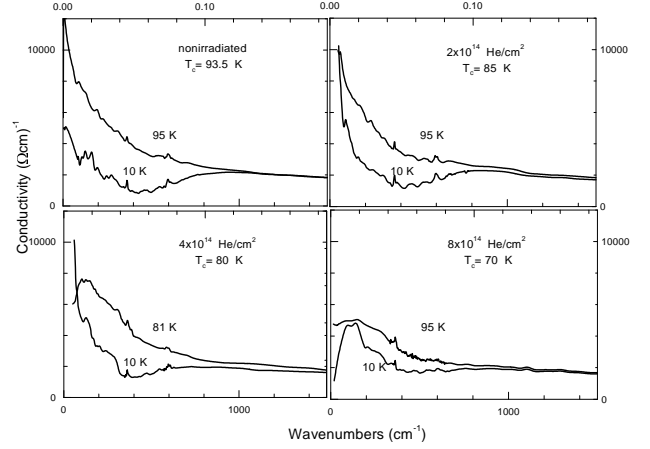}}
\caption{The effect of impurities on a d-wave superconductor. 
Increased defect concentration does not yield a gap signature in 
the conductivity as seen in BCS superconductors.  As the damage 
level is increased a Drude-like peak develops at low frequency. 
At high damage levels the peak moves to finite frequency. 
} \end{figure}

Fig 7 shows the effect of defects on the optical conductivity of radiation 
damaged YBa$_2$Cu$_3$O$_{7-\delta}$ from the work of Basov {\it et 
al.}\cite{basov94b} 
As the defect concentration rises, a 
Drude-like peak, displaced slightly from zero frequency, grows. There is no 
sign of a peak at $2\Delta$, expected to be in the neighborhood of 600 
cm$^{-1}$. This behavior is seen in a number of cuprate systems containing 
either doped impurities\cite{basov98} or other forms of disorder 
such as partial occupancy at oxygen sites\cite{xue93,timusk95} or radiation 
damage.\cite{basov94b} Recent work of Basov {\it et al.} shows 
similar effects in Zn doped YBa$_2$Cu$_3$O$_{7-\delta}$: there is 
no $2\Delta$ gap signature 
and, in the superconducting state there is a Drude-like conductivity peak, 
shifted slightly from zero to finite frequencies.\cite{basov98} A BCS s-wave 
superconductor is relatively insensitive to disorder where the main result 
of strong impurity scattering is the erasure of any gap 
anisotropy and the removal of the clean limit restriction on 
momentum conservation resulting in 
the familiar $2\Delta$ gap signature. In contrast, in a d-wave superconductor 
impurity scattering has the effect of restoring part of the Fermi surface 
near the nodes and the optical conductivity acquires a low frequency 
component, very similar to a Drude peak.\cite{carbotte94} 


\begin{figure}[t]
\leavevmode
\epsfxsize=\columnwidth
\centerline{\epsffile{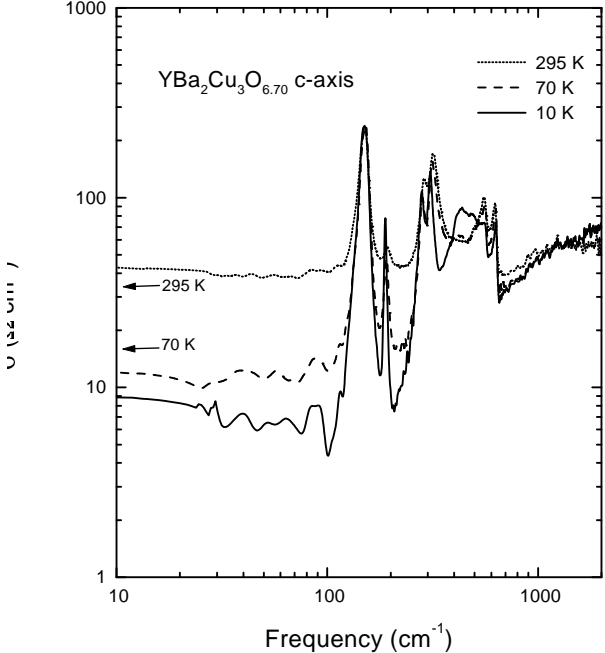}}
 \caption{Conductivity normal to the copper oxygen planes of 
YBa$_2$Cu$_3$O$_{7-\delta}$. Over two orders of magnitude in 
frequency, the electronic background conductivity is frequency 
independent. The peaks are due to optic phonons and the 
depression in conductivity below 300 cm$^{-1}$ is due to the 
pseudogap. } \end{figure}

The second fundamental property of the cuprates is an intrinsic anisotropy of 
electronic transport that goes beyond the simple anisotropic effective 
mass seen for example in doped semiconductors or in the Sr$_2$RuO$_4$ system 
discussed above. Here we have a qualitative difference in the properties 
along the planes where the conductivity is coherent and metallic and 
perpendicular to the planes, in the c-direction, where the conductivity is 
incoherent down to the lowest temperatures and frequencies. Fig 
8 shows the 
c-axis conductivity of a YBa$_2$Cu$_3$O$_{7-\delta}$ crystal that is 
slightly underdoped.  Like the Sr$_2$RuO$_4$ discussed in the last section, 
at high temperature there is a broad, essentially frequency independent, 
absorption band and no coherent Drude peak.  Since the material shown is 
underdoped it shows a depressed conductivity below 300 cm$^{-1}$ due to the 
normal state pseudogap at low temperature\cite{homes93} that 
will be discussed below. But down to the 
lowest frequency and temperature, there is no evidence of a Drude peak. The 
arrows denote the values of the dc resistivity and it clear that, at least 
above 70 K, there is no coherent conduction in this system. It is not 
possible to measure the dc resistivity below this temperature since the 
material becomes superconducting.  Recent microwave 
measurements\cite{hosseini98} show that the conductivity remains incoherent 
down to 22 GHz and 1.2 K. 
                               

\begin{figure}[t]
\leavevmode
\epsfxsize=\columnwidth
\centerline{\epsffile{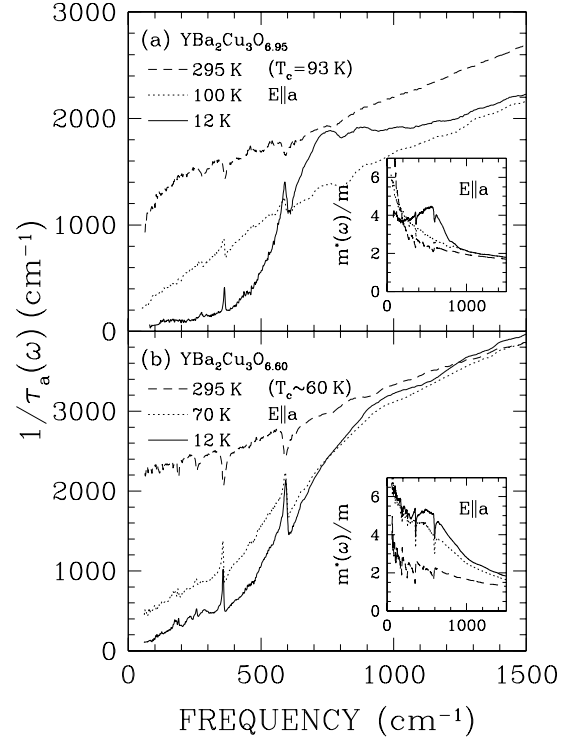}}
\caption{Frequency dependent scattering rate of the holes in 
YBCO, a) optimally doped sample and b) an underdoped sample.  
In the optimally doped sample the scattering rate is linear in 
both frequency and temperature above $T_c$ and develops a gap 
like depression below $T_c$. In the underdoped sample the 
scattering rate depression sets in well above $T_c$ which is 
evidence of a normal state pseudogap. The insets show the 
effective mass of the carriers which show a characteristic 
resonance at 600 cm$^{-1}$. } \end{figure}

Recent work using a variety of experimental probes, most strikingly angle 
resolved photoemission,\cite{loeser96,ding96}  shows that the normal state 
of the high temperature superconductors is dominated by a partial gap, or a 
pseudogap. The experimental evidence has been reviewed recently by Timusk 
and Statt.\cite{timusk99} The pseudogap forms at a temperature $T^*$ which 
is substantially  above the superconducting transition temperature in most 
underdoped samples. $T^*$  approaches $T_c$ in YBCO near 
optimal doping. Just as infrared spectroscopy can show the collapse of 
scattering associated with the development of superconductivity, it also 
shows the loss of electron-electron scattering on formation of the 
pseudogap. As shown first by Basov {\it et al.}\cite{basov96c} and 
illustrated in Fig. 9, from a recent paper of Homes {\it et 
al.}\cite{homes99} there is a loss of ab plane scattering associated with 
the formation of the pseudogap for frequencies below about 660 cm$^{-1}$. 

Fig 8 shows a depression in the conductivity associated with the pseudogap that 
develops in this underdoped sample. There is a gaplike region of flat 
conductivity that fills in as the temperature is increased but there appears to 
be no change in the overall frequency scale of the conductivity as the 
temperature is raised -- the gap fills in rather than closing. These 
measurements of the c-axis pseudogap by Homes {\it et al.}\cite{homes93} were 
the  first to yield spectroscopic evidence of the pseudogap. The c-axis 
pseudogap has since been observed in several other HTSC systems where samples 
of sufficient thickness are available\cite{basov94g,reedyk97} as well as in the ladder 
cuprates.\cite{osafune99}

In summary, infrared spectroscopy has revealed several 
exotic properties of the cuprates, the electronic coupling  mechanism, 
possibly associated with magnetism, the intrinsic anisotropy and finally the 
presence of a pseudogap in the normal state. In what follows we will see 
that organic low-dimensional superconductors share some of these properties 
and that they are the properties that can be used to define exotic 
superconductivity. 

\subsection{THE ORGANIC SUPERCONDUCTORS}

There are two main families of organic superconductors: conductors 
based on the two-dimensional (BEDT-TTF) organic molecule with a $T_c$ in the 
neighborhood of 10 K, and the family of one-dimensional materials, the 
Bechgaard salts, based on the TMTSF molecule with a $T_c$ just over 1 K. 

There is surprisingly little known about the electronic properties of 
organic conductors largely because of the small size and fragility of the 
available crystals. Thus magnetic neutron scattering and angle resolved 
photoemission, powerful, momentum sensitive techniques, have not proven to be 
as useful here as  they have been in the cuprates.  While there 
is a 
considerable body of infrared spectroscopy, the strong mid-infrared 
absorption seen in the organics by all investigators is in clear 
contradiction with the extensive low temperature magnetic transport data. 
These contradictions in the interpretation of the experiments have lead to 
two completely opposing views of the normal state transport particularly in the 
Bechgaard materials where most of the work has been done. 

Magnetic transport measurements have generally been interpreted in terms of 
Fermi liquid models with extraordinarily long scattering times. 
In agreement with this picture it is found 
that Kohler's rule holds\cite{forro84} except for the conductivity component 
along the conducting chains. Various oscillatory phenomena have been seen 
in high fields  and these have been interpreted in terms of electrons moving 
in quasi two-dimensional orbits. This interpretation of the data has been 
summarized by Greene and Chaikin\cite{greene84}. Specific heat 
data\cite{garoche82} also suggest a superconducting transition from 
a normal metallic state.

The infrared community takes a diametrically opposed stand on the transport 
properties. Measurements of the frequency dependent conductivity 
at low temperature shows that the very large dc conductivity of 
$\approx 200,000$ 
($\Omega$cm)$^{-1}$ retains its large value well into the 
microwave region\cite{gorshunov,dressel96} but then drops 
dramatically in  the 300 GHz range to a low value of $\approx 
1000$ ($\Omega$cm)$^{-1}$.\cite{tanner74} The reflectance of such a system 
is expected to show a prominent plasma edge in the 300 GHz 
region, a difficult experimental region to work in with small 
needle like samples. Nevertheless, early far infrared work of Tanner's group 
as well as recent backward wave oscillator data from Gr\"uner's group 
has shown evidence of this edge 
experimentally.\cite{tanner74,dressel96} 
A clear observation of this edge confirms the picture of two component 
conductivity, a narrow Drude peak, presumably caused by a 
density wave followed by  a very broad incoherent band due to a 
strongly correlated Luttinger liquid. 


\begin{figure}[t]
\leavevmode
\epsfxsize=\columnwidth
\centerline{\epsffile{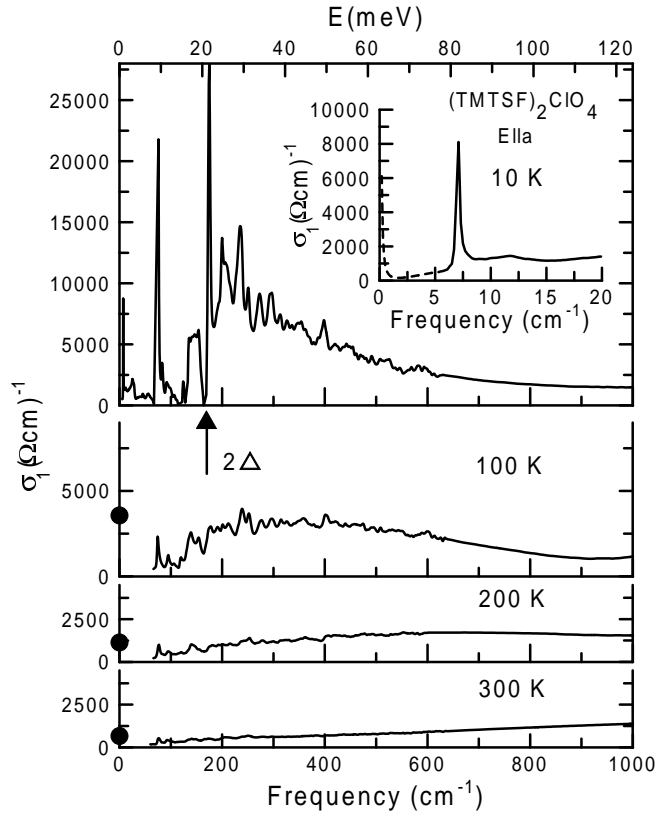}}
\caption{Real part of  $\sigma(\omega)$  of (TMTSF)$_2$ClO$_4$ 
along the chain axis. The full circles on the left axis are 
the values of the dc conductivity.  The 10 K dc 
value ($\sim2\times10^5$ $\Omega^{-1}{\rm cm}$ ) is too high to 
be shown on the scale used. A gap with the value $2\Delta\simeq 
170$ cm$^{-1}$\ can be seen at 10 K. The inset is $\sigma(\omega)$ 
at 10 K between 5 and 20 cm$^{-1}$.} 
  
\end{figure}

Fig. 10 shows recent conductivity data of (TMTSF)$_2$ClO$_4$ from 
Cao {\it et al.} for light polarized along the conducting 
chains.\cite{cao96} We see, at room temperature, a broad 
incoherent band very similar to  what is seen in the c-direction 
of the cuprates where the carriers are confined to the planes. 
The dc conductivity is not much larger than the infrared 
conductivity although there appears to be an incipient pseudogap 
already at room temperature.  

As the temperature is lowered the dc conductivity increases and spectral 
weight is shifted to lower frequencies. A clear, gap-like depression, can 
be seen below 300 cm$^{-1}$. This behavior has also been seen in the charge 
density wave system TTF-TCNQ\cite{basista90}. 

These data have been interpreted in terms of transport by charge density 
wave fluctuations.\cite{tanner74} Several characteristics point to this 
picture.  First, there is a growth in the intensity of phonon lines as the 
temperature is lowered, behavior that is usually associated with sliding 
density waves interacting with the lattice.\cite{rice76prl,rice77ssc} 
Secondly, estimates of the spectral weight associated with the narrow low 
frequency mode of $\approx 500 m_e$ suggest that phonons are involved as 
well. In this picture the peak in the far-infrared would correspond to 
direct transitions across the charge density wave gap\cite{lee}. 

The picture of transport along the chain axis by charge density wave 
fluctuations rests mainly on the assumption that strong infrared absorption 
is due to intrinsic bulk effects. It has been suggested that surface defects 
such as cracks may be responsible for the absorption and there is indeed 
scatter in the infrared data from various laboratories 
suggesting that 
sample quality may be a problem\cite{cao96}. However it should be noted that the simple 
picture of cracks predicts that the absorption should continue well into 
the microwave range as long as the penetration depth is smaller than the crack 
depth. 

There are several ways of investigating this problem. One of 
course is the use of samples with better controlled surfaces to 
see if the strong mid-infrared absorption is intrinsic. Assuming 
that it is, then reflectance measurements in high magnetic 
fields should be made to determine if sufficiently strong fields 
will break up the density waves and turn the materials into 
quasi one dimensional metals. There is preliminary evidence that 
this might be the case.\cite{ng83} 

Because of the low transition temperature and the low frequency range of the 
expected superconducting gap, little is known about the optical 
properties of the organic conductors in the superconducting 
state. It is hoped however, that when larger crystals become 
available, far infrared and microwave techniques will be used to 
investigate the issue of the superconducting gap as well as the 
transition region between the high dc and microwave conductivity 
and the strongly absorbing pseudogap range. 

\begin{figure}[t]
\leavevmode
\epsfxsize=\columnwidth
\centerline{\epsffile{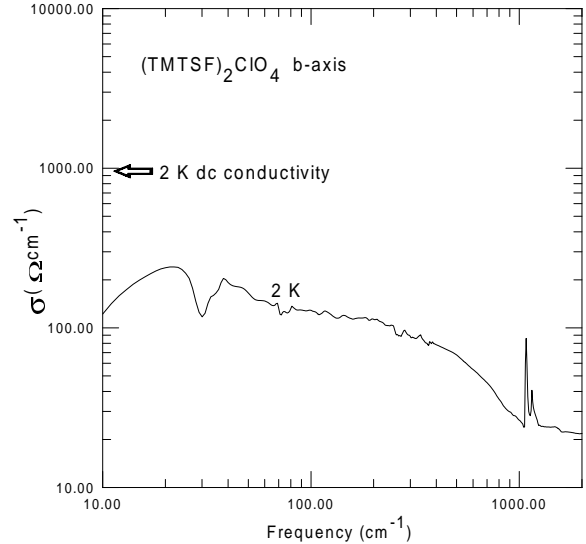}}
\caption{Conductivity normal to the chain direction in (TMTSF)$_2$ClO$_4$. The 
conductivity is flat and frequency independent. Its low value and 
lack of coherence suggests that the transport is incoherent normal to the 
conducting chains.
}
\end{figure}

We finally touch briefly on the question of transport normal to 
the conducting chains. Fig 11 shows the infrared conductivity of 
(TMTSF)$_2$ClO$_4$ in the  b 
direction where there is considerable overlap between the chains 
(the third direction has even lower coupling between the 
chains). We again see a flat conductivity, similar to the case 
of the c-axis cuprates.  There is some evidence of a collective 
mode since the dc conductivity is considerably higher than the 
infrared conductivity. However, the discrepancy is much smaller 
than what is observed in the chain direction. If one were to 
assign a scattering rate to this flat conductivity it would be 
of the order of 500 cm-1 (40 meV). This is to be compared with 
the estimated $t_{\perp}=40$ meV, the transfer matrix 
element normal to the chains. In terms of a Fermi liquid 
picture, one would then expect that below a crossover 
temperature of $T=40$ meV $=460$ K coherence would develop and 
the material would become a 2D Fermi liquid. It is clear that 
down to 2 K there is completely incoherent transport between the chains. 

In summary the organic superconductors display very anomalous properties in 
the infrared.  There is no evidence of simple metallic transport in any 
direction.  Along the chains the currents are carried by collective modes, 
possibly sliding charge density waves while normal to the chains there is 
complete incoherence similar to what is seen in interplane transport in the 
cuprates.
 
\subsection{CONCLUSIONS}

We have discussed in this review an infrared view of the transport 
properties of a variety of superconductors, from the conventional 
BCS strong coupling superconductor lead, to the exotic cuprates and organic 
charge transfer materials. The focus has been on several markers of deviation 
from classic phonon-coupled systems: 1) the collapse of 
scattering at $T_c$, a signature of an electronic mechanism 
since phonon scattering is unaffected, to first order, by the 
formation of a gap in the electronic system, 2) the growth of a 
Drude peak in the superconducting state of dirty samples and 
absence of a $2\Delta$ gap feature, both signatures of an 
order parameter with nodes 3) the  role of intrinsic 
dimensionality that goes beyond simple anisotropy where we find 
that the carriers are confined to planes and chains and interplane/chain 
conductivity is incoherent down to the lowest temperatures and 
frequencies, 4) the presence of a pseudogap in the normal state 
associated with charge transport by complex objects. 

For a true classification of all superconductors into anomalous 
and exotic varieties, as opposed to conventional ones, one needs 
data from all experimental techniques. Nevertheless the infrared 
is a good initial tool for this task since it can be applied to 
a wide range of materials being relatively modest in the demands it 
places on the crystal growers, who are of course responsible for 
much of the past progress in our understanding of unconventional 
superconductivity. 

Acknowledgements.

I would first like to acknowledge Georgorgios Varelogiannis for 
the suggestion to undertake this survey of the infrared 
properties  of unconventional superconductors and Jules Carbotte 
for valuable discussions on all aspects of superconductivity. 
The results described in this paper come in most part from the 
work of D.N.~Basov, D.B.~Bonn, N.~Cao, B.F.~Farnworth, 
C.C.~Homes, H.K.~Ng, R.~Hughes, J.J.~McGuire, A.~Puchkov, 
M.~Reedyk, T.~R\~o\~om T.~Strach and T.~Startseva at McMaster. I 
thank Y.~Tokura for permission to use Fig. 5. The crystal 
growers responsible for these results are K.~Bechgaard, D. 
Colson, 
B. Dabrowski, S.~Doyle, P.~Fournier, J.D.~Garrett, P.D.~Han, 
A.M.~Hermann, N. T.~Kimura, K.~Kishio, N.N.~Koleshnikov, 
H.A.~Mook, M.~Okuya, R.~Liang, W.D.~Mosley and D.A.~Payne. To them the 
experimentalists owe much.

\newpage

\end{document}